\documentclass[conference]{IEEEtran}
\IEEEoverridecommandlockouts
\usepackage{cite}
\usepackage{amsmath,amssymb,amsfonts}
\usepackage{algorithmic}
\usepackage{graphicx}
\usepackage{textcomp}
\usepackage{xcolor}
\usepackage{algorithm}
\usepackage{algorithmic}
\usepackage{booktabs}
\usepackage{caption}
\usepackage{array}
\usepackage{multirow}
\usepackage{colortbl}
\usepackage{pgfplots}
\pgfplotsset{compat=1.18}
\usepackage{tikz}
\usepackage{pgfplotstable}
\usetikzlibrary{patterns, shapes, arrows.meta}

\def\BibTeX{{\rm B\kern-.05em{\sc i\kern-.025em b}\kern-.08em
    T\kern-.1667em\lower.7ex\hbox{E}\kern-.125emX}}

\begin{document}

\title{MCBA: A Matroid Constraint-Based Approach for Composite Service Recommendation Considering Compatibility and Diversity
\\
(Invited Paper)

}

\author{\IEEEauthorblockN{1\textsuperscript{st} SunYing}
\IEEEauthorblockA{\textit{dept. name of organization (of Aff.)} \\
\textit{name of organization (of Aff.)}\\
City, Country \\
email address or ORCID}
\and
\IEEEauthorblockN{2\textsuperscript{nd} Given Name Surname}
\IEEEauthorblockA{\textit{dept. name of organization (of Aff.)} \\
\textit{name of organization (of Aff.)}\\
City, Country \\
email address or ORCID}
\and
\IEEEauthorblockN{3\textsuperscript{rd} Given Name Surname}
\IEEEauthorblockA{\textit{dept. name of organization (of Aff.)} \\
\textit{name of organization (of Aff.)}\\
City, Country \\
email address or ORCID}
}

\author{
    \IEEEauthorblockN{
        Ying Sun\IEEEauthorrefmark{1},
        Xiao Wang\IEEEauthorrefmark{2},
        Hanchuan Xu\IEEEauthorrefmark{2},
        Zhongjie Wang\IEEEauthorrefmark{2}
    }
    \IEEEauthorblockA{
        \textit{Faculty of Computing, Harbin Institute of Technology, Harbin, China} \\
        \IEEEauthorrefmark{1}sunying155644@163.com,
        \IEEEauthorrefmark{2}\{wxlxq, xhc, rainy\}@hit.edu.cn
    }
}

\maketitle

\begin{abstract}
With the growing popularity of microservices, many companies are encapsulating their business processes as Web APIs for remote invocation. These lightweight Web APIs offer mashup developers an efficient way to achieve complex functionalities without starting from scratch. However, this also presents challenges, such as the concentration of developers' search results on popular APIs limiting diversity, and difficulties in verifying API compatibility. A method is needed to recommend diverse compostions of compatible APIs that fulfill mashup functional requirements from a large pool of candidate APIs. To tackle this issue, this paper introduces a Matroid Constraint-Based Approach (MCBA) for composite service recommendation, consisting of two stages: API composition discovery focusing on compatibility and top-k composition recommendation focusing on diversity. In the first stage, the API composition issue is formulated as a minimal group Steiner tree (MGST) problem, subsequently addressed by a ``compression-solution" algorithm. In the second stage, a Maximum Marginal Relevance method under partition matroid constraints (MMR-PMC) is employed to ensure recommendation diversity. Comprehensive experiments on the real-world dataset show that MCBA surpasses several state-of-the-art methods in terms of accuracy, compatibility, diversity, and efficiency.

\end{abstract}

\section{Introduction}
Since the advent of Service-Oriented Architecture (SOA) \cite{b1}, software development has gradually shifted towards a ``user-centric'' model, promoting the creation of applications through Web services \cite{b2}. As the notions of ``servicization" and ``microservices" evolve, numerous services and applications have been encapsulated in lightweight API interfaces by various organizations, making them accessible online for mashup developers. As a way of integrating services, mashups combine one or more APIs into a composite application to meet developers' complex needs. By utilizing APIs, developers can avoid starting from scratch, significantly improving the efficiency of mashup service development \cite{b3}.

To better leverage the vast and functionally complex APIs, numerous researchers have proposed API recommendation methods specifically for mashup development \cite{b9,b10,r3}. These methods have achieved commendable results in uncovering API functionalities and recommending accurate APIs. Although these methods can save developers the time of manually searching for suitable APIs, they often overlook the compatibility checks, which may lead to the need for complex data format conversions during development. On the other hand, these methods prioritize the functionality and popularity of APIs in the recommendation process, resulting in recommendations that tend to focus on a subset of popular APIs. The lack of diversity not only limits developers' exploration of different functionalities and services but also hinders innovation.

To address the limitations of these methods, this paper introduces a Matroid Constraint-Based approach (MCBA) for composite service recommendation. MCBA initially constructs an API association graph based on historical mashup data to capture compatibility between APIs, which is then compressed into a supergraph. On this supergraph, dynamic programming technique is employed to identify connected subgraphs that cover the functional keywords in the developer's query. The connectivity of these subgraphs ensures that the obtained subgraphs are compatible API compositions. Subsequently, MCBA utilizes a quality-aware K-means algorithm to cluster these candidate API compositions into a partition matroid. Finally, MCBA implements a local greedy search strategy based on the MMR method, generating a diverse recommendation list while satisfying the constraints of the previously obtained partition matroid.
By simultaneously considering compatibility and diversity, MCBA is capable of providing users with efficient personalized mashup development support.

The main contributions of this paper are as follows:

\begin{itemize}
\item MCBA formulates the API composition challenge as a minimal group Steiner tree problem and innovatively introduces a ``compression-solution" algorithm to address the keyword search issue within this context. This addresses the issue of API compatibility while being highly efficient.
\item MCBA proposes MMR-PMC to ensure the diversity of recommendations. This approach combines the diversity capture methods of ``recommended items coming from different categories" and ``highest diversity scores", at the same time provides the advantage of ``controllable diversity".
\item Comprehensive experiments on the real-world dataset demonstrate that compared with several state-of-the-art methods, this method performs excellently, providing developers with more diverse, comprehensive, and efficient compatible composite service recommendations under similar accuracy and quality requirements.
\end{itemize}

The remainder of this paper is organized as follows: Section II reviews the related work. Section III defines the problem and introduces the theoretical background of MGST, Matroid Constraint, and MMR algorithms. The details of the MCBA are covered in Section IV, followed by the experiments in Section V. Finally, Section VI concludes the paper.

\section{Related Work}

With the widespread application of API technology, researchers from various fields have been actively exploring the use of APIs in mashup creation. Initial research primarily centered on enhancing mashup development via visualization tools \cite{r1,r2}. Subsequently, the researchers shifted their attention to the accuracy of the recommendations. Alshangiti et al. \cite{r3} introduced a novel Bayesian learning approach for API suggestions in mashup creation. Based on the neural graph collaborative filtering technique, Lian et al. \cite{r4} proposed an API recommendation method that exploits the higher-order connectivity between APIs and API users to improve recommendation accuracy. Li et al. \cite{r5} propose the web services recommendation based on metapath-guided graph attention network model to fully exploit the structural information of the knowledge graph. Despite their precision, these methods overlook a crucial aspect: compatibility between APIs.

Good compatibility relationships can help developers quickly develop without cumbersome data structure conversions and ensure the smooth operation of mashups. Yao et al. \cite{r6} model Web APIs compatibility as a hidden factor of descriptive similarities indicated by API profiles. Through the co-use of APIs in mobile APPs, the compatibility or complementary relations among APIs are introduced for better APIs recommendation. Wu et al. \cite{r7} model web APIs’ functions, popularity, and compatibility with an API correlation graph, then a top-k strategy is adopted in the recommendation process to guarantee popularity and compatibility. However, these methods often recommend API compositions that share many APIs. Given that their purpose is to provide optional API compositions for mashup developers, this lack of diversity significantly reduces the practicality of these methods.

To improve recommendation diversity, \cite{r8} introduced sampling techniques, but at the expense of success rate performance. Wang et al. \cite{r9} integrates a weighting mechanism and neighborhood information into matrix factorization to implement diversified and personalized APIs recommendations. Yu et al. \cite{r10} proposed a long-tail service recommendation method based on cross-view and comparison learning to enhance the diversity of recommendation results. While these methods improve diversity, they often sacrifice accuracy.

Therefore, there is a lack of a API recommendation approach that can simultaneously consider accuracy, compatibility, and diversity.

\section{Preliminaries}

\subsection{Problem Formulation}

The API recommendation task studied in this paper can be defined as follows: Given a graph $G$ and a functional requirement described by a sequence of keywords $Q = \{k_1, k_2, \dots, k_n\}$ (where each $k_i$ refers to a single keyword), the goal is to build a model to generate a diverse API composition recommendation list $T^{*} = \{T_{\text{top}1}, T_{\text{top}2}, \dots, T_{\text{top}k}\}$ where each item $T_{\text{top}i}$ is a compatible API composition that meets the functional requirements. The term "compatible" is defined as the ability of two APIs can be used together in a mashup without causing technical conflicts or necessitating complex data format conversions. For practical purposes, we consider APIs that have been successfully integrated in existing mashups to be compatible.

To find compatible API compositions, we first define an API association graph that aims to capture the compatibility relationships between APIs, inspired by \cite{datadriven}.

\textbf{Definition 1 Node:} For each Web API, there is a corresponding node $v$. Each node $v$ contains one or more keywords $\{k_1, k_2, \dots, k_n\}$ representing the functionalities provided by the API.

\textbf{Definition 2 Edge:} For each pair of compatible API nodes $v_i,v_j$, there exists a directed edge $e=(v_i,v_j)$. If $v_j$ is a successor node of $v_i$, the edge is directed from $v_i$ to $v_j$; conversely, the edge is directed from $v_j$ to $v_i$. If both are true, the edge can be bidirectional.

\textbf{Definition 3 Graph:} An API Association Graph is represented as $G = (V, E)$, where $V$ denotes the set of nodes and $E$ denotes the set of edges.

The problem of finding an API composition is to identify a set of APIs that are not only compatible but also collectively fulfill the functional requirements specified by the keyword sequence. This implies that these APIs constitute a connected subgraph within the API association graph.

\subsection{Minimum Group Steiner Tree}

In this subsection, we first provide the definition of the Steiner tree \cite{b20} and then model the problem of finding compatible API compositions as solving the Minimum Group Steiner Tree problem. To keep the discussion consistent, it needs to be stated that the graph used in the minimum group Steiner tree definition is indeed the API association graph introduced earlier.

\textbf{Definition 4 Steiner Tree:} Given a graph $G = (V, E)$ and a node set $V' \subseteq V$, if a tree $T$ can cover all nodes in the set $V'$, then $T$ is called a Steiner tree in the graph $G$ with respect to $V'$.

\textbf{Definition 5 Group Steiner Tree:} Given a graph $G = (V, E)$ and a node set $V^* = \{V_1, V_2, \dots, V_k\}(V_i \subseteq V, 1 \leq i \leq K)$, if $T$ is a Steiner tree and for each node set $V_i$, the tree $T$ covers only one node from the set $V_i$, then $T$ is called a Group Steiner Tree.

\textbf{Definition 6 Minimum Group Steiner Tree:} Given a graph $G = (V, E)$, a node set $V' \subseteq V$ and $n ( n > 0)$ Group Steiner Trees $T^* = \{T_1, T_2, \dots, T_k\}$ of $V'$, if $T \in T^*$ satisfies $w(T) = \min\{w(T_i) |T_i \in T^* \}$, then $T$ is called the Minimum Group Steiner Tree of the graph, where $w(T)$ denotes the weight of the tree.

\begin{figure}[htbp]
\centerline{\includegraphics[width=0.4\textwidth]{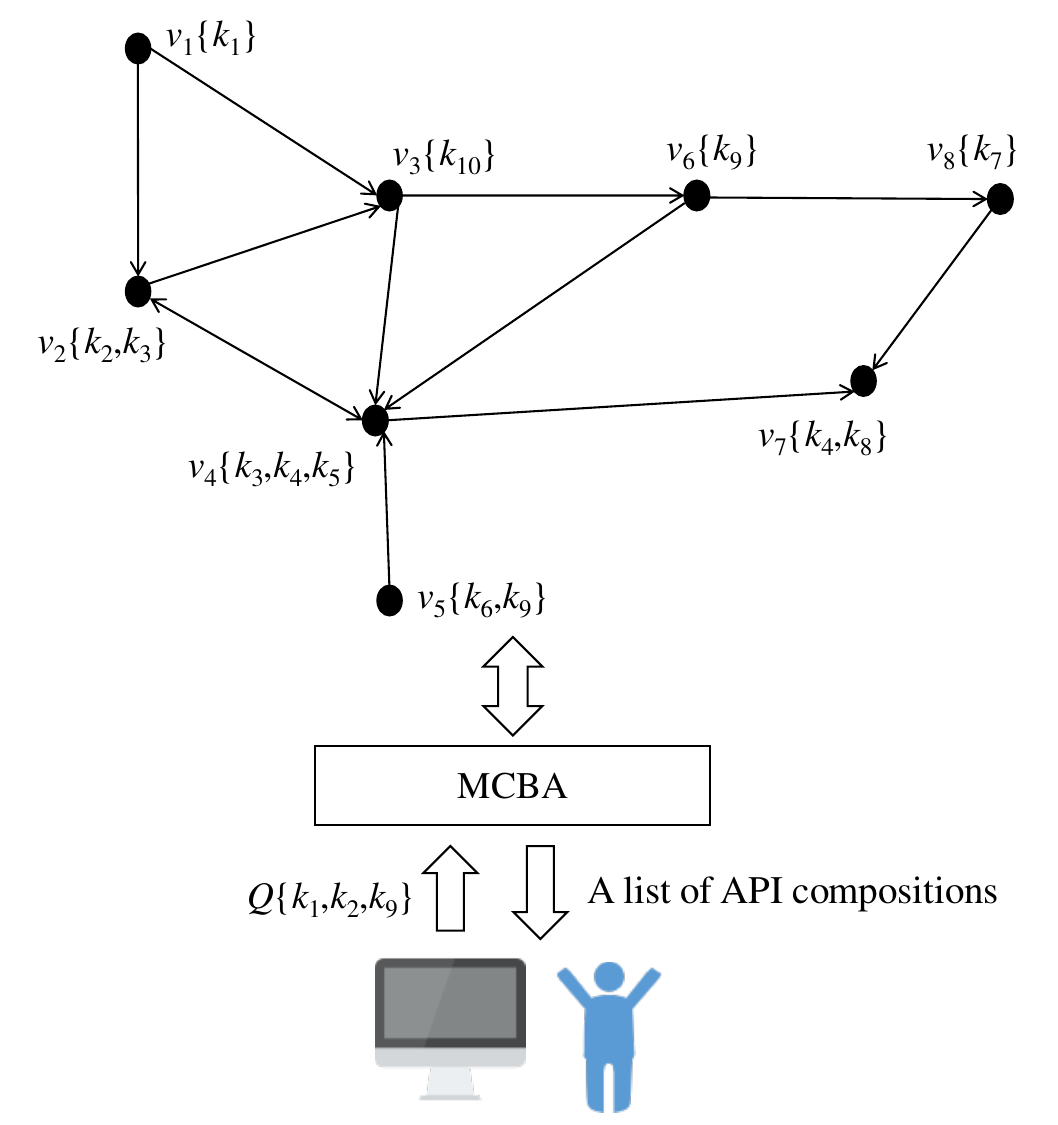}}
\caption{Example of API association graph.}
\label{model question}
\end{figure}

Based on the previously provided definition of an API association graph, the problem of finding API compositions can be modeled as the Minimum Group Steiner Tree problem. Taking Figure \ref{model question} as an example, the graph is generated from the following eight historical mashup-API development records.
\begin{align*} 
R_1&:\{v_1, v_3, v_4\}  &R_2&:\{v_4,v_2,v_3\} \\ 
R_3&:\{v_1,v_2\}        &R_4&:\{v_2,v_4,v_7\} \\ 
R_5&:\{v_5,v_4\}        &R_6&:\{v_4,v_7\} \\ 
R_7&:\{v_6,v_4\}        &R_8&:\{v_3,v_6,v_8,v_7\}
\end{align*}
So there are 8 nodes (APIs), namely $v_1,\dots,v_8$, covering functional keywords $k_1,\dots,k_{10}$. Symbol $v_1\{k_i,k_j\}$ denotes that API node $v_1$ can provide functionalities $k_i$ and $k_j$.

In Figure \ref{model question}, a mashup developer's functional requirements are represented by three keywords $Q\{k_1, k_2, k_9\}$. To achieve the required functionalities, we need to select APIs that cover these functionalities. As shown in Figure, node set $\{v_1\}$ covers keyword $k_1$, node set $\{v_2\}$ covers keyword $k_2$, and node set $\{v_5, v_6\}$ cover keyword $k_9$. Therefore, to meet the query, a tree needs to be found connecting one node from each of $\{v_1\}, \{v_2\}$ and $\{v_5, v_6\}$. However, since these three node sets are not directly connected, nodes that do not contain any keywords from $Q$ (i.e., bridging nodes) like $v_3$ need to be used. Therefore, the tree satisfying the functional requirements is actually a Group Steiner Tree \cite{b6}. Further, among the Group Steiner Trees satisfying the functional requirements, we aim to provide developers with the API composition that has the fewest APIs and the highest compatibility to reduce development costs, which is the Minimum Group Steiner Tree.

\begin{figure*}[h]
    \centering
    \includegraphics[width=1\textwidth]{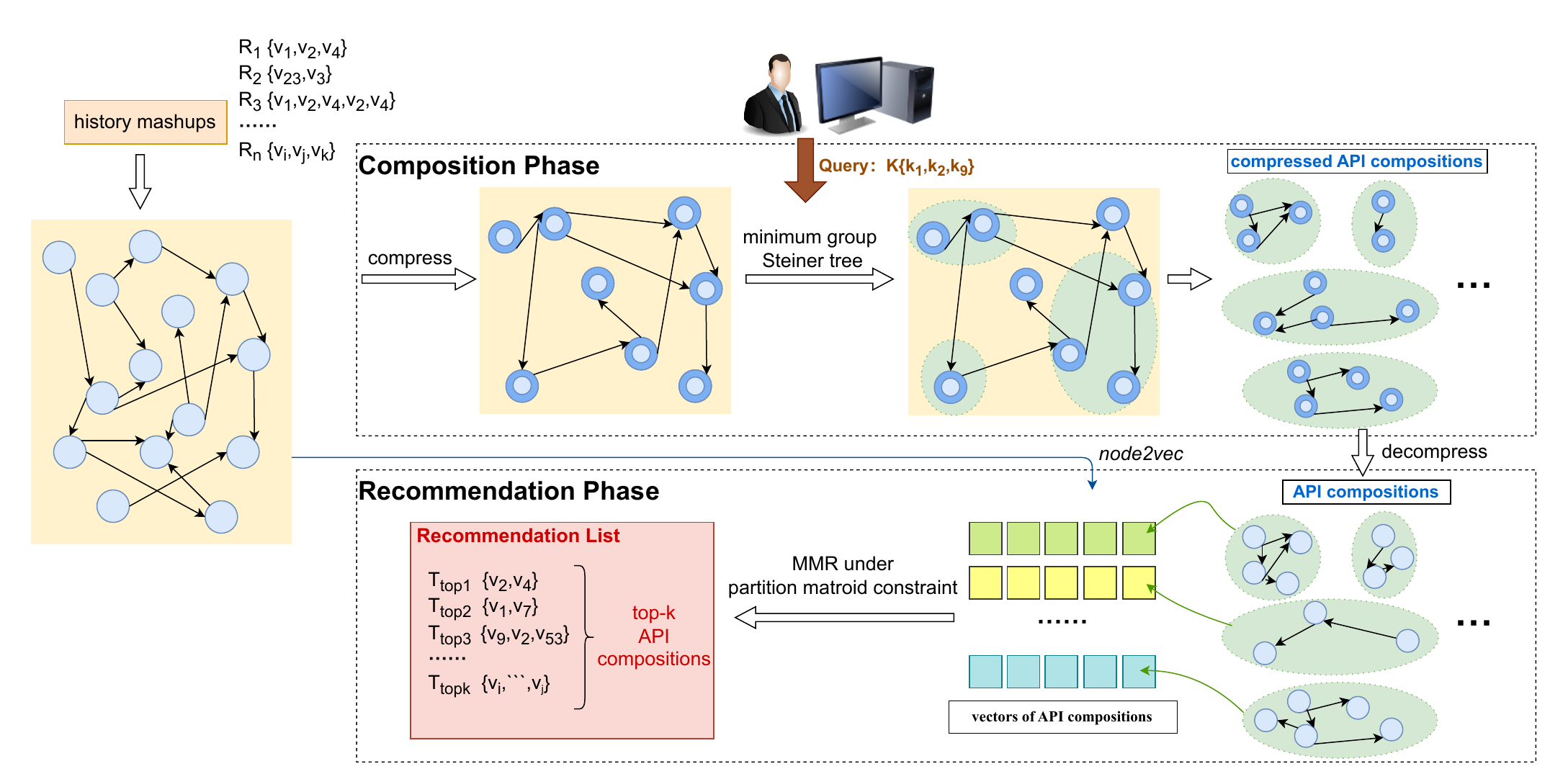}
    \caption{Illustration of MCBA framework.}
    \label{MCBA}
\end{figure*}

\subsection{Matroid Constraints and Maximum Marginal Relevance}

To achieve the goal of diversity, this paper proposes the MMR-PMC method, which combines two common ways of capturing diversity \cite{b21_twoways} during the recommendation phase. The relevant knowledge is presented here.

First, we focus on the first method of capturing diversity: the partition matroid constraint model. Matroid constraints \cite{b22_matroid} are a type of constraint widely used in optimization problems. For example, in graph theory and combinatorial optimization, matroid constraints can be used to describe some structural restrictions (such as minimum spanning trees) to help us find the optimal solution of these structures. Partition matroid constraints are a special type of matroid constraints suitable for the problem in this paper. 

\textbf{Definition 7 Matroid:} Given a finite non-empty set $E$ and a collection $\mathcal{I}$ of subsets of $E$, $(E, \mathcal{I})$ is a matroid if the collection satisfies the following three properties:
\begin{enumerate}
    \item  $\emptyset \in \mathcal{I}$.
    \item  If $A \in \mathcal{I}$ and $B \subseteq A$, then $B \in \mathcal{I}$.
    \item If $A, B \in \mathcal{I}$ and $|A| > |B|$, then there exists $e \in A \setminus B$ such that $B \cup \{e\} \in \mathcal{I}$.
\end{enumerate}

The collection $\mathcal{I}$ is the independent set family of $E$, and $ S \in \mathcal{I}$ is independent set.

\textbf{Definition 8 Partition Matroid:} Given a matroid $(E, \mathcal{I})$ and a partition $E^*=\{E_1, E_2, \dots, E_k\}(E_i \cap E_j = \emptyset, 1 \leq i \leq k)$ of all elements in $E$. $(E, \mathcal{I})$ is a partition matroid if and only if each independent set $ S \in \mathcal{I}$ contains at most one element from any $E_i$, i.e., $\forall S \in \mathcal{I}, |S \cap E_i| ( \leq 1 \leq i \leq k)$. At this time, the independent set $S$ satisfies the partition matroid constraint under $(E, \mathcal{I})$ and $E^*$.

\textbf{Motivation for using partition matroid constraints:} To achieve diversity in the recommendation list, a common approach is to require that items in the recommendation list come from different categories. Take the music site's ``Recommended Playlist" for example, all songs can be divided into different categories such as $\{\text{pop, folk, classical, etc.}\}$ , if we mandate that no two songs in the recommended playlist come from the same category, the result is naturally more diverse than the list without category restrictions. This constraint of ``not having elements from the same category in the recommendation list" can be viewed as a special matroid constraint—a partition matroid constraint.

Next, we focus on the second method of capturing diversity: maximizing the diversity score. While maximizing diversity, the high quality of the recommendation list should not be completely ignored. Therefore we introduce the Maximal Marginal Relevance (MMR) algorithm. MMR \cite{b23_mmr} is a method widely used in set-based information retrieval, which can simultaneously consider diversity and relevance during diversified recommendations.

Referring to \cite{b24, b25}, MMR algorithm is presented in a more general form. Suppose there is an item set $I$, and we aim to recommend a subset $S^*_q (S^*_q \subset I, |S^*_q|=q)$ related to a given query $Q$. MMR suggests building $S^*_q$ in a greedy way as in Formula \ref{mmr}:
when $S^*_k=\{s_1,\dots,s_k\}$ is already been chosen, Formula \ref{mmr score} is used to select $s_{k+1}$ to obtain $S^*_{k+1}(S^*_{k+1} = S^*_k \cap s_{k+1})$:
\begin{equation}
s_{k+1} = \arg\max_{s_i \in I \setminus S^*_k}\mathcal F (s_i, S^*_k)
\label{mmr}
\end{equation}
\begin{equation}
F (s_i, S^*_k)=\lambda\text{Sim}_1(s_i, Q)+(1-\lambda)\max_{s_j \in S^*_k}(\text{Sim}_2(s_i, s_j))
\label{mmr score}
\end{equation}

In the above formulas, $\text{Sim}_1(s_i, Q)$ measures the relevance between item $s_i$ and the query $Q$, $\text{Sim}_2(s_i, s_j)$ measures the distance between two items $s_i$ and $s_J$, and $\lambda$ is a parameter representing the relevance weight. It is clear that the selection mechanism used by the MMR algorithm during recommendations considers both relevance and diversity.

Combining the two methods of capturing diversity, we propose MMR-PMC method, using the MMR algorithm to select the top-k items that meet the partition matroid constraint from the candidate items as the final recommendation list. In MCBA, ``items" are instantiated as ``API compositions". 

\section{Matroid Constraint-Based Approach for Composite Service Recommendation}

This section will focus on describing in detail how MCBA works to get a diverse, compatible, and high-quality recommendation list of API compositions. Overall, the algorithm can be divided into two phases: composition discovery phase, and composition recommendation phase as in Figure \ref{MCBA}. In the first phase, the minimum Steiner tree is solved using the compression-solution approach to obtain a large number of compatible API compositions; in the second phase, the MMR-PMC method is used to filter the results of the first phase and recommend high-quality, diverse top-k API compositions.

\subsection{Composition Discovery Phase}

Solving the minimum group Steiner tree is an NP-complete problem \cite{b26}. This paper chooses to use the dynamic programming idea to gradually obtain the global optimal solution.

In the process of finding compatible API compositions that meet the functional requirements of developers, the data density of the graph is an influential factor to consider \cite{datadriven}. A high ratio of nodes to keywords within the graph structure can lead to a dispersion of APIs, rendering their interconnections more challenging due to the sparsity of functional keywords. On the other hand, a high ratio of edges to keywords within the graph can proliferate the number of potential subproblem solutions. This proliferation can inversely impact the efficiency of the dynamic programming algorithm, leading to a marked decrease in its performance. Under such conditions, the algorithm may fail to yield a viable solution within the stipulated timeframe.

\begin{figure}[h]
    \centering
    \includegraphics[width=0.4\textwidth]{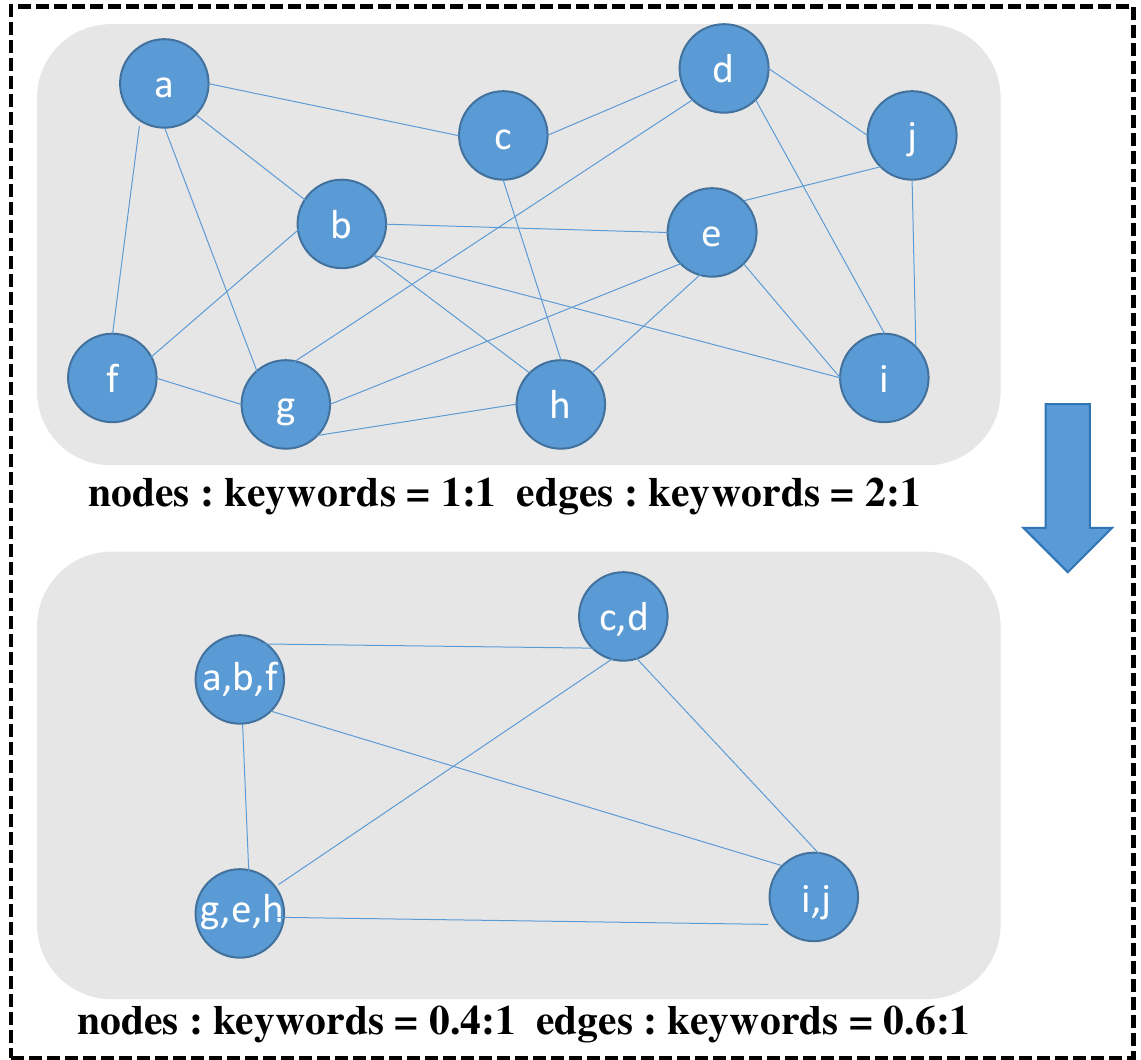}
    \caption{Schematic diagram of compression effect.}
    \label{fig:compression_effect}
\end{figure}

To cope with different data scales of the API association graph, MCBA proposes a ``compression-solution" algorithm. By compressing the original API association graph, we can obtain a compressed graph with uniform distribution of edges, nodes and keywords (e.g., Fig. \ref{fig:compression_effect}). This compression graph can significantly improve the success rate and time efficiency of solving the minimal group Steiner tree problem.

\subsubsection{\textbf{Step One: Compress the Original API Association Graph into SuperGragh}}

``Graph Compression" \cite{b27, b28} is a technique designed to simplify large-scale graphs by creating a condensed graph. This is achieved by merging multiple nodes into ``supernodes" or removing non-essential edges, reducing the graph's complexity while preserving its structural features, and facilitating subsequent analytical tasks like graph pattern mining and neighborhood queries \cite{b28}. We use ``reachability-preserving graph compression" \cite{b28} to ensure the minimum group Steiner tree from the compressed graph reflects the original connectivity. The relevant definitions are as follows:

\textbf{Definition 9 SuperNode:} A supernode $sv=(A,K)$ aggregates one or more ancestor nodes $A=\{v_1,\dots.v_k\}$ , with its weight being the number of ancestors. It inherits the functional keywords $K=\cup_{v_i\in A}\text{Keywords}(v_i)$ and adjacency relationships of its ancestors, ensuring connectivity among the ancestor nodes.

\textbf{Definition 10 SuperEdge:} A directed superedge $sv_i$ , $sv_j$ exists between supernodes $sv_i$ , $sv_j$, if there is an edge in the original graph from an ancestor of $sv_i$ to an ancestor of $sv_j$.

\textbf{Definition 11 SuperGraph:} The compressed API association graph is represented as $SG=(G,SV,SE)$, where $G$ is the original graph, $SV$ is the set of supernodes, and $SE$ is the set of superedges.

The pseudocode of the compression algorithm is as Algorithm \ref{code 1}. Given the original API association graph \(G\) and the compression granularity \(p\) (representing that a supernode can have up to \(p\) ancestor nodes), the supernode set of the compressed graph is first established (lines 3-15): starting with any original node, adjacent nodes are added in turn to form a set of original nodes with a number less than \(p\). Then all the nodes in this set are aggregated into a supernode. This process continues until all original nodes are processed. Finally, all edges are updated (lines 16-19) by updating both ends of an edge supernodes instead of the original nodes, thus obtaining the supergraph. 

\begin{algorithm}
\caption{Reachability-Preserving Graph Compression Algorithm}
\begin{algorithmic}[1]
\algsetup{linenodelimiter=}
\REQUIRE Original API association graph $G=(V,E)$, compression granularity $p$
\ENSURE Supergraph $SG$
\STATE Initialize an empty set of supernodes ${SV}$
\STATE Initialize an empty set of superedges ${SE}$
\FOR{each node $v \in G$}
    \IF {$v$ hasn't been used yet}
        \STATE Initialize a set $S=\{v\}$
        \STATE $U$  is the set of adjacent nodes of $S$
        \WHILE{$|S| < p$ and $U\neq \emptyset$}
            \STATE Select a neighbor $u$ from $U$
            \STATE Add $u$ to $S$
            \STATE Update $U$
       \ENDWHILE
    \STATE Create a supernode $sv$ from $S$
    \STATE Add $sv$ to $SV$
    \ENDIF
\ENDFOR
\FOR{each edge $(u, v) \in E$}
    \STATE Create a superedge $se$ from $(u,v)$
    \STATE add $se$ to $SE$
\ENDFOR
\STATE $SG \leftarrow (G, SV, SE)$
\STATE \textbf{return} $SG$
\end{algorithmic}
\label{code 1}
\end{algorithm}

\subsubsection{\textbf{Step Two: Solve the Minimal Group Steiner Tree Problem Using Dynamic Programming}}

Given a query's set of functional keywords $K$, the goal is to find the minimal group Steiner tree $ST(sv,K)$ on the compressed supergraph $SG$. Here, $sv$ is the root supernode.

Using dynamic programming, we break down this problem into manageable subproblems, solving each and storing the results to build the optimal solution iteratively. More specifically, a minimum group Steiner tree $ST(sv,K)$ with a specified height $h$, which is the maximum distance from the root to any leaf can be found by successively enlarging group Steiner trees of incremental heights $h=0,1,\dots$, covering the keywords in $K' \subseteq K$ \cite{b29}.

The state $ST(sv,K')$ in the dynamic programming framework has a weight $w(ST(sv,K'))$ , representing the number of web APIs in the tree, equal to the sum of the weights of all supernodes. The state transition equations are:

\begin{equation}
w(ST_{min}(sv,K'))=w(sv)  IF |K'|=1
\label{dynamic 1}
\end{equation}
\begin{align}
 &w(ST_{min}(sv,K'))= \notag \\
 &min(w(ST_g(sv,K)), 
             w(ST_m(sv,K'))) 
\label{dynamic 2}   
\end{align}
\begin{align}
&w(ST_g(sv,K'))=  \notag \\
&\min_{su \in N(sv)}{w(ST_{min}(su,K')+w(sv))}
\label{dynamic 3}    
\end{align}
\begin{align}
 &w(ST_m(sv,K'))= \notag \\ 
 &\min_{\substack{K'=K'_1 \cup K'_2 \\ K'_1 \cap K'_2 =\emptyset}}\{w(ST_{min}(sv,K'_1) \oplus ST_{min}(sv,K'_2) )\}
\label{dynamic 4}
\end{align}

The set $N(sv)$ represents the collection of adjacent nodes to the supernode $sv$ in $SG$. Equation \ref{dynamic 1} indicates that a tree covering one keyword has the weight of the associated supernode. Equation \ref{dynamic 2} demonstrates that the tree $ST_{min}(sv,K')$ can be obtained by either growing the tree ( \ref{dynamic 3} ) or merging trees ( \ref{dynamic 4} ). Growing involves adding $sv$ to a tree rooted at a neighbor $su$ , while merging combines two trees rooted at $sv$ covering disjoint subsets $K'_1$ and  $K'_2$.

These processes of growth and merging continue until the tree fully covers the keywords set $K$ , resulting in the final minimum group Steiner tree, which represents a compatible API composition fulfilling the developer's functional needs. Algorithm \ref{code 2} is the pseudo-code for the algorithm.

\begin{algorithm}
\caption{The minimal group Steiner tree algorithm}
\begin{algorithmic}[1]
\algsetup{linenodelimiter=}
\REQUIRE $SG=(G, SV, SE)$: The supregraph after compression ; $F=\{f_1, f_2, \dots, f_n\}$: A collection of functional keywords set for each node $sv_i(1<i<n)$ ; $K=\{k_1, k_2, \ldots, k_l\}$: Collection of functional keywords in query
\ENSURE $ST^*=\{ST_1,ST_2,\dots,ST_L\}$: A collection of minimal group Steiner Trees where each tree is a compatible APIs composition that covers the functional keywords $K$ 

\STATE Initialize $Q$ and $ST^*$ according to the weights of the trees. The queue $Q$ is sorted in ascending order.
\STATE $Q \gets \emptyset$, $ST^* \gets \emptyset$
\FOR {each $sv \in SV$}
    \STATE $K' \gets f_{sv} \cap K$
    \IF {$K' \neq K$}
        \STATE enqueue $ST_{\min}(sv, K')$ into $Q$
    \ENDIF
\ENDFOR
\WHILE {$Q \neq \emptyset$}
    \STATE dequeue $Q$ to $ST_{\min}(sv, K')$
    \IF {$K' = K$}
        \STATE add $ST_{\min}(sv, K')$ into $ST^*$
        \STATE \textbf{continue}
    \ELSE
        \FOR {each $su \in N(sv)$}
            \IF {$w(sv) + w(ST_{\min}(su, K')) < w(ST_{\min}(sv, K'))$}
                \STATE $ST_{\min}(sv, K') = se(su, sv) + ST_{\min}(su, K')$
                \STATE enqueue $ST_{\min}(sv, K')$ into $Q$
                \STATE update $Q$
            \ENDIF
        \ENDFOR
        \STATE $K_1' \gets K'$
        \FOR {each $ST_{\min}(sv, K_2')$ in $Q$}
            \IF {$K_1' \cap K_2' = \emptyset$}
                \IF {$w(ST_{\min}(sv, K_1')) + $ $w(ST_{\min}(sv, K_2')) < w(ST_{\min}(sv, K_1' \cup K_2'))$}
                    \STATE $ST_{\min}(sv, K_1' \cup K_2') = ST_{\min}(sv, K_1') \oplus ST_{\min}(sv, K_2')$
                    \STATE enqueue $ST_{\min}(sv, K_1' \cup K_2')$ into $Q$
                    \STATE update $Q$
                \ENDIF
            \ENDIF
        \ENDFOR
    \ENDIF
\ENDWHILE
\STATE \RETURN $ST^*$
\end{algorithmic}

\label{code 2}
\end{algorithm}

In Algorithm \ref{code 2}, a priority queue $Q$ is maintained to store the generated trees sorted by their node counts. Initially, for each supernode 
$sv \in SV$ with a non-empty keyword $K' \cap K \neq \emptyset$, the corresponding minimum tree $ST_{\min}(sv,K')$ is enqueued (lines 3-8). Subsequently, the algorithm then dequeues tree $ST_{\min}(sv,K')$ from $Q$ and checks if it covers all keywords in $K$. If so, it is added to the result list (lines 11-13). If not, the tree is either grown by adding a neighbor supernode (lines 16-19) or merged with another tree (lines 22-28), and the new tree is re-enqueued. This continues until $Q$ is empty.

The final output is a list of the minimal group Steiner trees, each covering all keywords from the query $Q$. However, since these trees are composed of supernodes, they must be decompressed to obtain a list of API composition candidates suitable for the recommendation phase.

\subsubsection{\textbf{Step Three: Decompress All the Trees}}

In this step, we decompress the supernodes in the minimal group Steiner trees obtained earlier to filter out redundant APIs. Each supernode is decomposed into its ancestor API nodes, ensuring that the resulting node set forms a connected subgraph in the original API association graph due to the ``reachability-preserving graph compression" algorithm. We then identify and remove two types of useless nodes: those without any useful keywords and those with duplicate keywords. This process maintains connectivity and yields a list of candidate API compositions without redundant nodes.

Decompressing all the minimal group Steiner trees provides us with a list of candidate API compositions, which serves as the input for the subsequent phase.

\subsection{Composition Recommendation Phase}

In the previous phase, we obtained a candidate list of compatible API compositions, and the current goal is to select a diverse top-k list using the MMR-PMC method. Recalling Section \uppercase\expandafter{\romannumeral3}, the method involves clustering all API compositions to form a partitioned matroid, and then applying the MMR algorithm under the partitioned matroid constraints defined by the cluster.

Considering our recommendation goals and algorithmic needs, we utilize historical mashup data to evaluate the quality of API compositions. We believe thatpast mashup developers favored high-quality APIs and pairs with high joint quality. Thus, an API composition's quality is determined by the quality of its individual APIs and the quality of their associations, as defined by Formula \ref{quality}, where $Times(v_i)$ refers to the number of times $v_i$ has been used historically, and $Times(v_i,v_{i+1})$ refers to the number of times the API pair $(v_i,v_j)$ has been historically co-invoked.

\begin{align}
Quality(T)=& \frac{\sum_{0 \leq i \leq n}Times(v_i)}{|T|}\notag\\
&+ \frac{\sum_{0 \leq i \leq n-1}Times(v_i,v_{i+1})}{|T|^2} 
\label{quality}
\end{align}

To assess the similarity between API compositions, we employ the inner product of feature vectors as the metric. Given the notable success of graph embedding techniques in recommendation systems \cite{b30,b31}, we introduce the $node2vec$ method \cite{b32} developed by Stanford University, which captures network structure and node neighborhood information, outputting a low-dimensional feature vector for each API node. Given $T=\{v_1,v_2,\dots,v_n\}$, $Vector(v_i)$ denotes the feature value of the API node $v_i$ in the same dimension. We then calculate the feature vector for each API composition using Formula \ref{vector} and normalize it with Formula \ref{norvector}. The similarity between two API compositions $U$ and $V$ is determined by Formula \ref{similarity}. The greater its value, the more similar $U$ and $V$ are.

\begin{equation}
    Vector(T)= Average(\sum_{1 \leq i \leq n}Vector(v_i))
\label{vector}
\end{equation}
\begin{equation}
    NorVector(T)=\frac{Vector(T)}{|Vector(T)|}
\label{norvector}
\end{equation}

\begin{align}
&Similarity(U,V)=  \notag \\
&\frac{NorVector(U) \times NorVector(V)}{||NorVector(U)|| \times ||NorVector(V)||}
\label{similarity}
\end{align}

It is important to note that the graph embedding operation was performed on the uncompressed API correlation graph to preserve the graph's characteristics. 

\subsubsection{\textbf{Step Four: Cluster to Obtain a Partitioned Matroid}}

We partition the list of API compositions into clusters to establish the partition matroid constraints for the next step. To simultaneously consider quality and diversity, we use a ``quality-aware K-means" approach. Unlike standard K-means, which selects initial cluster centers randomly, leading to inconsistent clustering \cite{b33}, this method initializes centers based on the quality of API compositions, then iterates until convergence. This process ensures a partition matroid with a balanced quality distribution.

\subsubsection{\textbf{Step Five: MMR-PMC Method}}

In this final step, we apply a local greedy search strategy \cite{b34} to implement the  MMR-PMC method derived from the previous step. The ultimate goal is to generate a recommendation list that is compatible, diverse, and high-quality. Algorithm \ref{code 3} outlines this process:

More specifically, the Algorithm \ref{code 3} begins by selecting a set of API compositions with the highest quality to form the initial recommendation list (lines 4-10). Then, it traverses the candidate list of API compositions in descending order of quality (lines 11-23). For each pair of API compositions $(T_\text{in},T_\text{out})$ that are either in or not in the current recommendation list, if the MMR score (which will be explained in detail later) of $T_\text{out}$ is higher than the score of $T_\text{in}$ (line 16), $T_\text{in}$ should be removed from the list and $T_\text{out}$ is added. After completing the loop, a top-k recommendation list is generated. Crucially, when attempting to add a new API composition, the algorithm ensures that it does not violate the partition matroid constraint by preventing two candidates from the same cluster from being in the list (lines 6 and 15). If this condition is not met, the swap is abandoned. In this way, the top-k list obtained by the MMR method ultimately satisfies the constraints of the partition matroid, capturing the diversity of the recommendation list from two dimensions. The pseudo-code for the Algorithm \ref{code 3} is as follows:

\begin{algorithm}
\caption{Greedy Search Algorithm under Partition Matroid Constraint.}
\algsetup{linenodelimiter=}
\begin{algorithmic}[1]
\REQUIRE $T^* = \{T_1, T_2, \ldots, T_n\}$: A list of API composition candidates; $d^* = \{d_1, d_2, \ldots, d_n\}$: Cluster number to which each API composition belongs; $k$: Number of API compositions to be recommended; $\lambda$: Relevance parameter
\ENSURE $T_{\text{top}}^* = \{T_{\text{top}1}, T_{\text{top}2}, \ldots, T_{\text{top}k}\}$: A list of top-k recommendations for API compositions

\STATE $T_{\text{top}}^* = \emptyset$, $D_{\text{top}} = \emptyset$
\STATE \textbf{sort} $T^*$ by $Quality(T_i)$ $(1 \leq i \leq n)$
\STATE $i = 0$
\WHILE {$|T_{\text{top}}^*| < k$}
    \STATE $i = i + 1$
    \IF {$d_{T_i} \notin D_{\text{top}}$}
        \STATE $T_{\text{top}}^* = T_{\text{top}}^* \cup \{T_i\}$
        \STATE $D_{\text{top}} = D_{\text{top}} \cup \{d_{T_i}\}$
    \ENDIF
\ENDWHILE

\FOR {each $T \in T^*$}
    \FOR {each $T_{\text{top}} \in T_{\text{top}}^*$}
        \STATE $T_{\text{top}}^{*'} = T_{\text{top}}^* - \{T_{\text{top}}\}$
        \STATE $D_{\text{top}}' = D_{\text{top}} - \{d_{T_{\text{top}}}\}$
        \IF {$d_T \notin D_{\text{top}}'$}
            \IF {$mmrScore(T,T_{\text{top}}^{*'}, \lambda) >mmrScore(T_{\text{top}}, T_{\text{top}}^{*'}, \lambda)$}
                \STATE $T_{\text{top}}^* = T_{\text{top}} \cup \{T\}$
                \STATE $D_{\text{top}} = D_{\text{top}}' \cup \{d_T\}$
            \STATE \textbf{continue}
            \ENDIF
        \ENDIF
    \ENDFOR
\ENDFOR

\RETURN $T_{\text{top}}^*$
\end{algorithmic}
\label{code 3}
\end{algorithm}

In Algorithm \ref{code 3}, $D_{\text{top}}$ is the set of clusters that $T^*_\text{top}$ have been currently covered. $mmrScore(T,T^*_\text{top},\lambda)$ \cite{b23_mmr} denotes the score of a candidate API composition $T$ under the current recommendation list $T^*_\text{top}$, which is defined by Formula \ref{api mmrscore}. 
\begin{align}
&mmrScore(T,T^*,\lambda)= \notag \\
&\lambda(Quality(T))- 
(1-\lambda)\max_{T_j \in T^*}(Similarity(T,T_i))
\label{api mmrscore}
\end{align}

 The plus sign in the original MMR formula has been replaced with a minus sign, as excessive similarity is considered a negative factor in diversity recommendation. This formula simultaneously considers the quality of the API composition and its similarity to the existing recommendations, where a higher $\lambda$ value emphasizes quality, and a lower $\lambda$ value emphasizes diversity. This flexibility enables the recommendation system to cater to different user preferences, whether they prioritize diversity or quality.

After completing the five steps outlined, we end up with a diverse list of recommendations. Each item in this list is a carefully selected API composition that not only meets the functional requirements of the users but also excels in terms of compatibility and quality.

\subsection{Time complexity analysis}

In the MCBA algorithm, Steps One, Two and Five are most time-consuming. Assume that the API association graph is $G=(V, E)$ in which $|V|=n$, $|E|=m$, the compressed graph is $(G, SV, SE)$ in which $|SV|=sn$, $|SE|=sm$; the set of functional keywords in the query is $K=\{k_1, k_2, \ldots, k_l\}(|K|=l)$; the size of the list of candidate API compositions obtained in "composition discovery phase" is $t$, and the size of recommendation list obtained in "composition recommendation phase" is $k$.

\textbf{Step One:} Compress the Original API Association Graph into SuperGragh. The graph compression algorithm involves merging nodes into supernodes and updating the pointers of edges, which takes \( O(n + m) \) time in the worst case, as each node and edge is visited once.

\textbf{Step Two:} The time complexity of solving the minimal group Steiner tree problem using dynamic programming is $O(2^l \cdot sn \cdot \log(2^l \cdot sn) + 2^l \cdot sm + 3^l \cdot sn)$ \cite{datadriven}. 

\textbf{Step Five:} In the MMR method, we first need to sort the candidate API composition list of size $t$, which has a time complexity of $O(t \log t)$ using the heap sort algorithm. Then, the algorithm iterates over the candidate list, comparing the score of each API composition against the current recommended list. Since there are
$k$ API compositions, the loop's time complexity is $O(tk)$. Consequently, the overall time complexity of Step Five is
$O(t\log t+tk)$, which combines the sorting of candidates and the loop's operations.

In summary, the time complexity of MCBA is:
 $O(n+m + 2^l \cdot sn \cdot \log(2^l \cdot sn) + 2^l \cdot sm + 3^l \cdot sn +t\log t+tk)$. Additionally as explained in section \uppercase\expandafter{\romannumeral5}, $l$ is usually never greater than 6, and $k$ is also usually a constant of a fixed size. So we can simplify it to $O(n+m+sn \log(sn) + sm +sn + t\log t +t)$. Further, it is clear that $O(n+sn)=O(n),O(m+sm)=O(m)$  according to the characteristics of the supergraph. Therefore, the final time complexity is ultimatly expressed as $O(n+m+sn \log(sn) +t\log t+t)$.

\section{Experiments}
In this section, we first delineate the experimental setup and then demonstrate the effectiveness of MCBA across three dimensions. \footnote{The code is available on GitHub: https://github.com/Felicity155/MCBA.git.}

\subsection{Experimental setup}
The experiments use a dataset \cite{b18_dataset} collected by Liu et al. from an API sharing platform, encompassing 7,767 mashups and 23,580 historical interactions between APIs. We consider that the first Category keyword of an API represents its primary function, serving as its functional keyword; the functional keywords of a mashup are the union of the functional keywords  of all APIs that compose it. Following this principle, a sequence of query is established. Statistical analysis revealed that the keyword sequences of mashups in dataset \cite{b18_dataset} ranged from 3 to 6, as is the length of the query sequence. The program was run on a computer with a Windows 11 system, an Intel(R) Core(TM) i5-1035G1 CPU @1.19GHz, and 16GB of RAM.

In order to measure the performance of the algorithm, the following experimental metrics are used. Indicators 6 and 7 are only used to measure the performance of the ``composition discovery" phase.
\begin{itemize}
\item \textbf{Mean Precision (MP) \cite{b37}:} Given a query and an API composition, we consider APIs that coincide with the real-world call history as ``correct" APIs. Precision is the proportion of correctly recommended APIs to all APIs in the composition. Mean Precision is the average precision of all query results, and the higher the value, the better the algorithm performs.
\item \textbf{Mean Quality (MQ):} Given an API composition, its quality is determined by the historical call frequency of individual APIs and the historical joint call frequency of API pairs, as defined in section \uppercase\expandafter{\romannumeral4}. Mean Quality is the average quality of all query results, and the higher the value, the better.
\item \textbf{Mean Inter-list Diversity (MID) \cite{b38}:} Given a recommendation list for a query, inter-list diversity is measured by the Hamming distance, which quantifies the dissimilarity between all pairs of API compositions. The Hamming distance is defined as Formula \ref{hamming}:
\begin{equation}
    H_{ij} = 1 - \frac{|T_{\text{top}i} \cap T_{\text{top}j}|}{|T_{\text{top}i}| + |T_{\text{top}j}|}
    \label{hamming}
\end{equation}
    Mean Inter-List Diversity is the average of inter-list diversity values across all queries, with higher values indicating greater diversity in the recommendations.
    \item \textbf{Coverage \cite{datadriven}:} It is the percentage of APIs that have appeared in the recommendation list out of the total APIs. Higher values are preferable, as diverse recommendation results should not be limited to popular APIs.
    \item \textbf{Time Cost (TC):} The time cost is one of the key indicators to measure the performance of the algorithm, and the smaller the value, the better.
    \item \textbf{Success Rate (SR) \cite{datadriven} :} Given a keyword query, if an API composition is found with a length less than twice that of the query, it is deemed a successful solution for developers. Success Rate is the proportion of queries that can be successfully resolved.
    \item \textbf{Mean Size (MS) \cite{b39}:} The mean size is the average length of all API compositions in the recommended list, with smaller values indicating more concise solutions.
\end{itemize}

\subsection{Parameter calibration}
First, we focus on how does the correlation parameter $\lambda$ affect the performance of the algorithm.

\begin{figure}[h]
    \centering
    \includegraphics[width=0.5\textwidth]{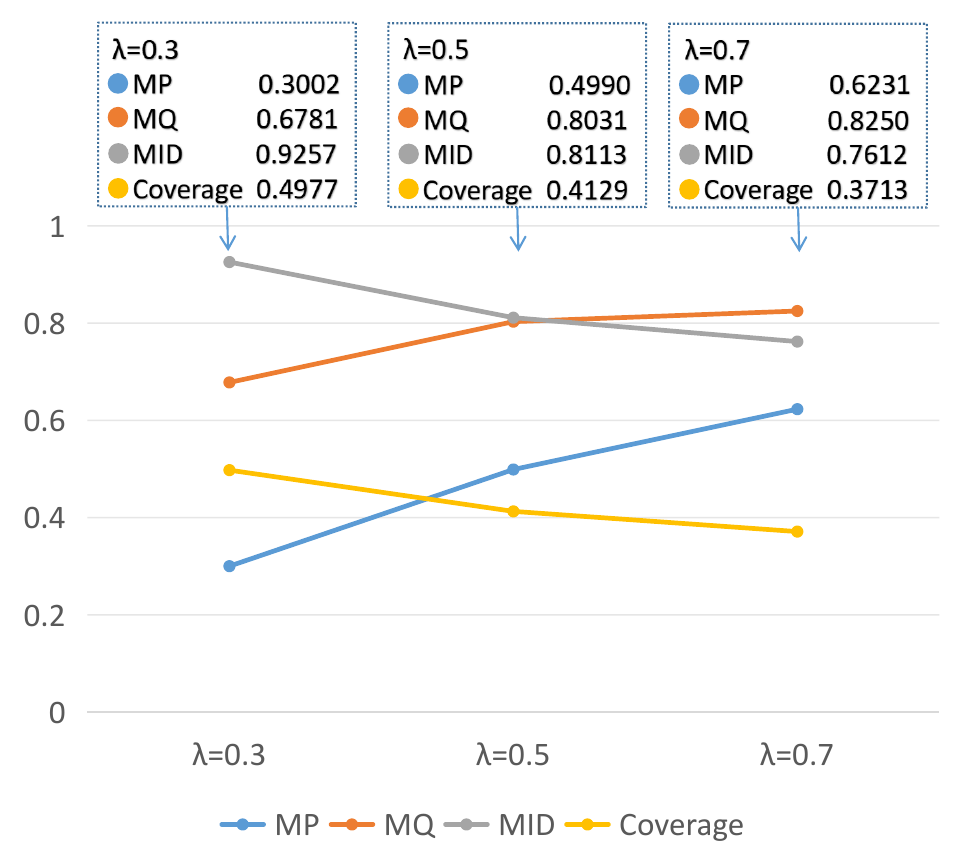}
    \caption{The performance of the algorithm under different $\lambda$}
    \label{lambdaTable}
\end{figure}

In the MMR method, $\lambda$ is a parameter that defines the importance of ``relevance". The larger $\lambda$ is, the greater the proportion of the quality factor in the API composition score, resulting in a more precise recommendation list; conversely, the smaller $\lambda$ is, the more diverse the recommendation list. Our goal is to simultaneously consider accuracy and diversity of the recommendation results by adjusting $\lambda$. By setting $\lambda$ to 0.3, 0.5, and 0.7, the following data in Figure \ref{lambdaTable} is obtained. To enhance the visual clarity of the graph, MQ is standardized the for uniformity. 

We can observe that as $\lambda$ increases, the MP and MQ significantly increase, while the MID and Coverage decrease to some extent. This shows that the method in this paper retains the ``controllable diversity" characteristic of MMR, allowing developers to freely adjust the characteristics of the recommendation list by specifying $\lambda$.

In Figure \ref{lambdaTable}, the MID and Coverage rate remain above 0.75 and 0.35, respectively, higher than the average diversity and coverage rate of other five methods in Table \ref{compare result1} and \ref{compare result2}: 0.7171 and 0.3323. This is because MCBA combines two diversity capture methods: partition matrix constraints and the MMR method. Even when MMR focuses more on relevance, partition matrix constraints still ensure that the items in the recommendation list are not too similar. In scenarios where higher precision and quality are required, the method can still guarantee the diversity of recommendations.

\subsection{Ablation experiment}
Next, we explore whether the ``compression-solution" algorithm improve the success rate and efficiency of API composition discovery compared to traditional dynamic programming approaches.

We conducted experiments on API association graphs with different data densities (node ratios of 1:5 and 1:10) to obtain 50 compatible API compositions that cover the query keywords. In this subsection, the goal is to investigate how the ``compression-solution" algorithm performs compared to the direct use of the dynamic programming algorithm. The results are presented in Table \ref{compression result1} and \ref{compression result2}.

\begin{table}[h]
\centering
\caption{On the graph with Nodes: Edges = 1:5}
\renewcommand{\arraystretch}{1.5} 
\begin{tabular}{
    @{}>{\centering\arraybackslash}p{1.5cm}@{\hspace{0cm}}
    >{\centering\arraybackslash}p{1cm}@{\hspace{0cm}}
    >{\centering\arraybackslash}p{1.4cm}@{\hspace{0cm}}
    >{\centering\arraybackslash}p{1cm}@{\hspace{0cm}}
    >{\centering\arraybackslash}p{1.4cm}@{\hspace{0cm}}
    >{\centering\arraybackslash}p{1cm}@{\hspace{0cm}}
    >{\centering\arraybackslash}p{1.4cm}@{}}
\toprule
\multicolumn{1}{c}{\multirow{2}{*}{\shortstack{Number\\of Queries}}} & \multicolumn{2}{c}{SR} & \multicolumn{2}{c}{MS} & \multicolumn{2}{c}{TC} \\ 
\cmidrule(r){2-3} \cmidrule(r){4-5} \cmidrule(r){6-7}
 & Comp & Non-comp & Comp & Non-comp & Comp & Non-comp \\ 
\midrule
347  & 1.00 & 1.00 & 4.02 & 3.49 & 3.90 & 20.02 \\ 
500  & 0.50 & 0.34 & 4.46 & 3.59 & 3.83 & 5.35 \\ 
2000 & 0.47 & 0.35 & 4.41 & 3.63 & 1.63 & 3.28 \\ 
\bottomrule
\end{tabular}
\label{compression result1}
\end{table}

\begin{table}[h]
\centering
\caption{On the graph with Nodes: Edges = 1:10}
\renewcommand{\arraystretch}{1.5} 
\begin{tabular}{
    @{}>{\centering\arraybackslash}p{1.5cm}@{\hspace{0cm}}
    >{\centering\arraybackslash}p{1cm}@{\hspace{0cm}}
    >{\centering\arraybackslash}p{1.4cm}@{\hspace{0cm}}
    >{\centering\arraybackslash}p{1cm}@{\hspace{0cm}}
    >{\centering\arraybackslash}p{1.4cm}@{\hspace{0cm}}
    >{\centering\arraybackslash}p{1cm}@{\hspace{0cm}}
    >{\centering\arraybackslash}p{1.4cm}@{}}
\toprule
\multicolumn{1}{c}{\multirow{2}{*}{\shortstack{Number\\of Queries}}} & \multicolumn{2}{c}{SR} & \multicolumn{2}{c}{MS} & \multicolumn{2}{c}{TC} \\ 
\cmidrule(r){2-3} \cmidrule(r){4-5} \cmidrule(r){6-7}
 & Comp & Non-comp & Comp & Non-comp & Comp & Non-comp \\ 
\midrule
347  & 1.00 & 1.00 & 3.78 & 3.46 & 7.96 & 752.82 \\ 
500  & 0.62 & 0.61 & 4.09 & 3.78 & 8.52 & 58.03 \\ 
2000 & 0.59 & 0.58 & 4.06 & 3.70 & 6.18 & 28.82 \\  
\bottomrule
\end{tabular}
\label{compression result2}
\end{table}

\begin{table*}[ht]
    \centering
    \caption{Performances of the methods in recommending top-5 API compositions}
    \renewcommand{\arraystretch}{2} 
    \begin{tabular}{@{}l*{5}{>{\centering\arraybackslash}p{1cm}}@{}}
        \toprule
        Method & MP & MQ & MID & Coverage & TC \\
        \midrule
        Greedy    & 0.3519 & 20.1753 & 0.7791 & 0.4001 & 19.2983 \\
        MSD       & 0.4207 & 25.8400 & 0.7002 & 0.3943 & 35.0044 \\
        ATD-JSC   & 0.4720 & 25.9237 & 0.7215 & 0.3461 & 30.0299 \\
        DAWAR     & \textbf{0.5183} & \textbf{33.4910} & 0.6900 & 0.2997 & 40.4023 \\
        SSR     & 0.1052 & 17.8321 & 0.6911 & 0.2211 & 381.2243 \\
          \rowcolor{lightgray!30}
        MCBA    & 0.4990 & 32.1245 & \textbf{0.8113} & \textbf{0.4192} & \textbf{4.2653} \\
        \bottomrule
    \end{tabular}
    \label{compare result1}
\end{table*}

\begin{table*}[ht]
    \centering
    \caption{Performances of the methods in recommending top-10 API compositions}
    \renewcommand{\arraystretch}{2} 
    \begin{tabular}{@{}l*{5}{>{\centering\arraybackslash}p{1cm}}@{}}
        \toprule
        Method & MP & MQ & MID & Coverage & TC \\
        \midrule
        Greedy    & 0.2576 & 19.2367 & 0.7920 & 0.4602 & 29.3781 \\
        MSD       & 0.4125 & 23.4909 & 0.8013 & 0.4864 & 32.3244 \\
        ATD-JSC   & 0.4338 & 20.0072 & 0.7523 & 0.4671 & 49.2345 \\
        DAWAR     & \textbf{0.4945} & 29.6102 & 0.7029 & 0.4136 & 55.9021 \\
        SSR     & 0.2062 & 22.3532 & 0.7221 & 0.2806 & 881.8352 \\ 
        \rowcolor{lightgray!30}
        MCBA & 0.4764 & \textbf{30.0004} & \textbf{0.8201} & \textbf{0.5011} & \textbf{7.6010} \\
        \bottomrule
    \end{tabular}
    \label{compare result2}
\end{table*}

With Table \ref{compression result1} and \ref{compression result2}, we can learn that the ``compression-solution" algorithm generally outperforms the direct dynamic programming algorithm in the API composition phase when solving the minimum group Steiner tree problem. It has demonstrated a significant average improvement of 21.18\% in success rate and 70.34\% in time efficiency, while the average solution size has only experienced a marginal increase of approximately 0.53 units. The compression operation allows many originally adjacent nodes to appear in the same tree without requiring growth and merging, thus enhancing the algorithm's efficiency. Consequently, some redundant API nodes may appear, which cannot be eliminated even after decompression. However, considering the substantial improvement in time efficiency, these slight redundancies are acceptable.

The time cost for both methods decreases as the number of queries increases, which seems counterintuitive. This phenomenon occurs because the average time cost for a set of queries is significantly influenced by a small number of queries that take a long time to process. When the number of queries increases, this effect is averaged over a larger number of queries, making the impact less noticeable. Therefore, the average time cost more accurately reflects the true situation as the number of queries increases.

To achieve the optimal comprehensive effect, we set the $\lambda$ parameter of the MCBA algorithm to 0.5  in the subsequent experiments.

\subsection{Comparative Analysis}
Finally, we looked at how does MCBA perform compared to existing state-of-the-art methods.

To observe the effectiveness of the proposed method, we selected three advanced methods, namely ATD-JSC \cite{b40_ATDJSC}, DAWAR \cite{b36_dawar}, and SSR \cite{b41_SSR}, as well as two baseline methods, Greedy \cite{b23_mmr} and MSD \cite{b42_msd} , for comparison. For fairness, the parameter settings of the five reference methods follow \cite{b40_ATDJSC} \cite{b36_dawar}\cite{b41_SSR}\cite{b23_mmr}\cite{b42_msd}, and the feature vector embedding dimension in MCBA is 128 \cite{b36_dawar}. The number of candidate API compositions for all methods is 500. The results are shown in Table \ref{compare result1} and Table \ref{compare result2}.
\begin{itemize}
    \item \textbf{ATD-JSC \cite{b40_ATDJSC}:} This approach builds a similarity graph encompassing all potential API compositions and subsequently identifies the top-k compositions by extracting the maximum independent set from this graph.
    \item \textbf{DAWAR \cite{b36_dawar}:} Utilizing the Determinantal Point Processes (DPP) framework, this model assesses both the quality and similarity of candidate API compositions through a kernel matrix. It then computes the submodular function maximization to generate a top-k recommendation list.
    \item \textbf{SSR \cite{b41_SSR}:} This technique introduces a bilinear model to gauge the functional relevance between mashup requirements and services in a supervised context. It further employs a novel hypergraph-based approach to evaluate the composition preference among services. The final recommendation list is composed of APIs with the highest utility function scores.
    \item \textbf{Greedy \cite{b23_mmr}:} This straightforward method begins by randomly selecting an initial API composition and proceeds to iteratively incorporate the least similar composition from the remaining candidates until the top-k list is complete.
    \item \textbf{MSD \cite{b42_msd}:} In contrast to the MMR discussed in our paper, MSD formalizes diversity as the cumulative distance between pairs of items. Essentially, its goal is to produce a set that maximizes the total distance between its elements.

\end{itemize}.

We can learn that overall MCBA exhibits superior performance compared to other methods. More specifically, MCBA outperforms other methods in terms of MQ, MID, Coverage, and TC while performing equally in MP, which is acceptable (refer to the next point). We attribute its outstanding performance to two main factors: (1) leveraging node2vec for extracting feature vectors of API compositions, thereby better exploring the structural characteristics of the API correlation graph, and (2) combining the MMR method with  partition matroid constraint, which integrates two diversity capturing approaches while simultaneously considering diversity and relevance.

Our method achieves comparable performance in terms of MP, as expected. In reality, the majority of historical call data is concentrated on a small subset of popular APIs, indicating that including more popular APIs results in higher precision. However, considering diversity, the recommendation list should not be limited to popular APIs. Thus, MCBA aims to cover as many APIs as possible, providing developers with more diverse, comprehensive, and efficient recommendation results while maintaining comparable precision.

Regarding time expenditure, our method incurs significantly lower costs compared to others, primarily due to the ``compression-solution" technique used in the first phase. This approach compresses nodes into super nodes, drastically reducing time-intensive tasks such as tree growth in dynamic programming. The time saved outweighs that spent on compression and decompression, enabling quicker delivery of recommendation services to developers.

\section{Conclusion}

The emergence of Service-Oriented Architecture (SOA) and microservices has revolutionized software development by facilitating the creation of applications through APIs, simplifying mashup development for complex requirements. However, current API recommendation methods often lack compatibility checks and focus solely on popular APIs, resulting in intricate integration issues and limited diversity in recommendations.  This paper presents a new approach: MCBA: A Matroid Constraint-Based Approach for
composite service recommendation to provide solutions for mashup developers. To ensure compatibility, MCBA first uses the compression-solution method to solve the minimum Steiner tree problem to obtain a large number of compatible API compositions. Then, the MMR-PMC method is used to recommend high-quality and diversified top-k API compositions. MCBA can help developers create mashups by finding a variety of compatible API compositions that meet functional requirements based on the keyword sequence entered by developers. This paper has demonstrated the huge advantages of MCBA in experiments on real-world dataset.

In the future, we plan to continue focusing on the ``cold start" issue of API mashups recommendations, trying to use specific information (Tags, Description, etc.) of API and mashup to further explore the role of non-popular APIs in mashups. On this basis, a more diversified and novel composite service recommendation can be realized.

\section*{Acknowledgments}
The research presented in this paper has been supported by the National Key Research and Development Program of China (No.~2021YFF0900900) and the Key Research and Development Program in Heilongjiang Province (2022ZX01A28).

\bibliographystyle{IEEEtran}
\bibliography{IEEEabrv,mcba}

\end{document}